%% file: _Main.tex
\documentclass[a4paper, 11pt, oneside]{article} 
\usepackage{geometry}
\usepackage{booktabs} 
\usepackage{mathptmx} 
\usepackage{fancyhdr}
\usepackage[normalem]{ulem}
\usepackage[utf8]{inputenc}
\usepackage{microtype}
\usepackage{graphicx}
\usepackage{colortbl}
\definecolor{mygray}{gray}{.88}
\usepackage{multirow}
\usepackage{amsmath}
\usepackage{bm}
\usepackage{epsfig}
\usepackage{authblk}
\usepackage{amssymb,enumitem}
\usepackage[hyphens]{url}
\usepackage{hyperref}
\usepackage{color}
\usepackage{soul}
\usepackage{bbding}
\definecolor{mygray}{gray}{.88}
\usepackage{graphicx, subfig}
\usepackage{listings}
\usepackage{fontawesome}
\usepackage{longtable}
\usepackage{algorithm}
\usepackage{algorithmic}
\usepackage{appendix}
\newcommand{\tabincell}[2]{\begin{tabular}{@{}#1@{}}#2\end{tabular}}

\begin{document} 

\begin{titlepage} 

	\centering 
	
	\scshape 
	
	\vspace*{\baselineskip} 
	
	
	\rule{\textwidth}{1.6pt}\vspace*{-\baselineskip}\vspace*{2pt} 
	\rule{\textwidth}{0.4pt} 
	
	\vspace{0.75\baselineskip} 
	
	{\LARGE ToL:\\ A Tensor of List-Based\\ Unified Computation Model} 
	
	\vspace{0.75\baselineskip} 
	
	\rule{\textwidth}{0.4pt}\vspace*{-\baselineskip}\vspace{3.2pt} 
	\rule{\textwidth}{1.6pt} 
	
	\vspace{2\baselineskip} 
	
	
	
	\vspace*{3\baselineskip} 
	
	
	Edited By
	
	\vspace{0.5\baselineskip} 
	
	{\scshape\Large Hongxiao Li\\ Wanling Gao\\ Lei Wang\\Jianfeng Zhan\\ }
	
	
	\vspace{0.5\baselineskip} 

	\vfill 
	
	
	\epsfig{file=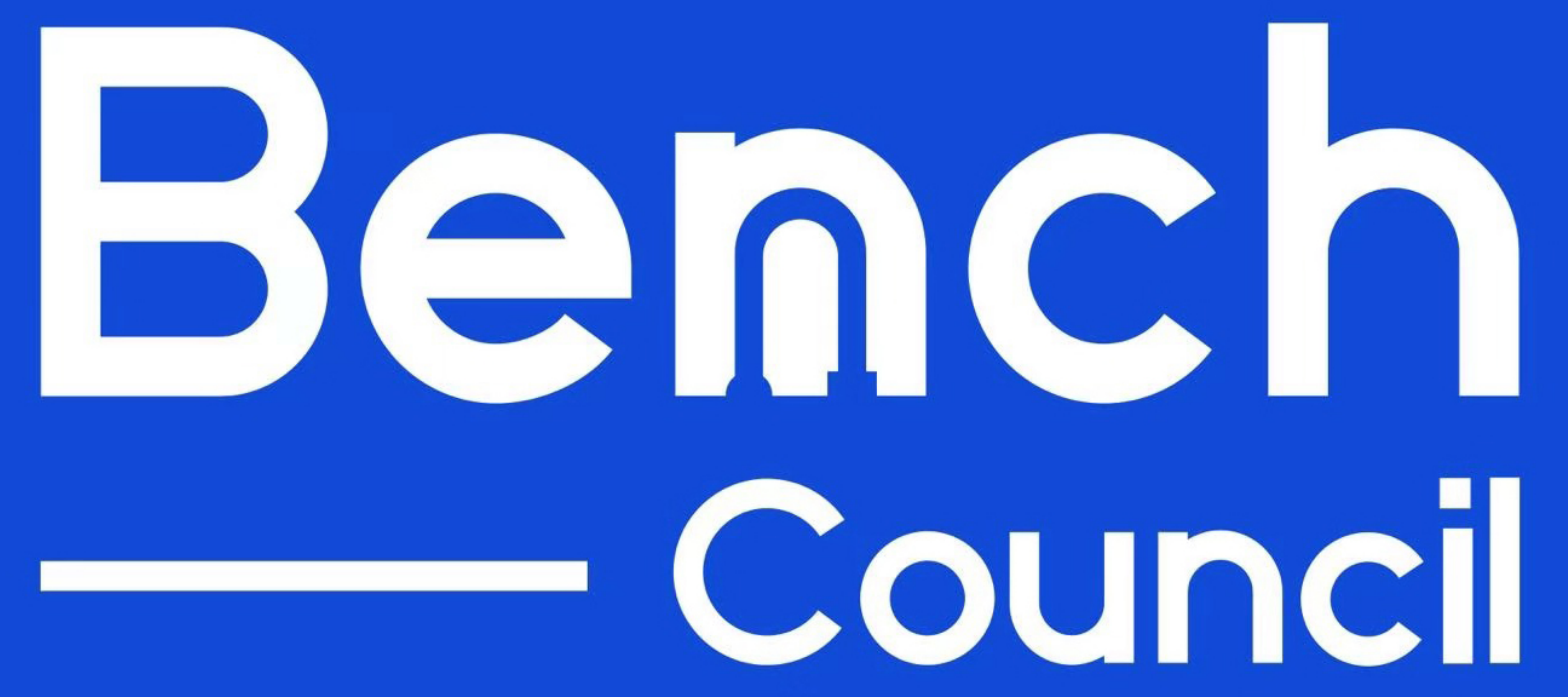,height=2cm}
	\textit{\\BenchCouncil: International Open Benchmark Council\\Chinese Academy of Sciences\\Beijing, China\\https://www.benchcouncil.org} 
	\vspace{5\baselineskip} 

	Technical Report No. BenchCouncil-ToL-2022 
	
	{\large Dec 20, 2022} 

\end{titlepage}


\title{ToL: A Tensor of List-Based Unified Computation Model}

\author[1,3]{Hongxiao Li}
\author[1,2,3]{Wanling Gao}
\author[1,2,3]{Lei Wang}
\author[1,2,3]{Jianfeng Zhan\thanks{Jianfeng Zhan is the corresponding author.}}

\affil[1]{Research Center for Advanced Computer Systems, State Key Lab of Processors, Institute of Computing Technology, Chinese Academy of Sciences\\ \{gaowanling, wanglei\_2011, zhanjianfeng\}@ict.ac.cn}
\affil[2]{BenchCouncil (International Open Benchmark Council)}
\affil[3]{University of Chinese Academy of Sciences\\ lihongxiao19@mails.ucas.ac.cn}

\date{Dec 20, 2022}
\maketitle

\begin{abstract}

Previous computation models either have equivalent abilities in representing all computations but fail to provide primitive operators for programming complex algorithms or lack generalized expression ability to represent newly-added computations. This article presents a unified computation model with generalized expression ability and a concise set of primitive operators for programming high-level algorithms. We propose a unified data abstraction -- Tensor of List, and offer a unified computation model based on Tensor of List, which we call the ToL model (in short, ToL). ToL introduces five atomic computations that can represent any elementary computation by finite composition, ensured with strict formal proof. Based on ToL, we design a pure-functional language -- ToLang. ToLang provides a concise set of primitive operators that can be used to program complex big data and AI algorithms. Our evaluations show ToL has generalized expression ability and a built-in performance indicator, born with a strictly defined computation metric -- elementary operation count (EOPs), consistent with FLOPs within a small error range.

\end{abstract}

\input{1.Introduction}
\input{2.Related}
\input{3.ToL}

\input{4.ToLang}

\input{5.Expression}

\input{6.Performance}
\input{7.Conclusion}

\bibliographystyle{ieeetr}
\bibliography{sample-bibliography}

\end{document}

%% file: 1.Introduction.tex
\section{Introduction}
Good data and computation abstractions catch the intrinsic properties of computing and hence play an essential role. The earliest computation models, such as Turing machine~\cite{turing1936computable}, Game of Life~\cite{adamatzky2010game}, Langton's ant~\cite{langton1986studying}, and lambda calculus~\cite{barendregt1984lambda}, are typical example.   These models have equivalent abilities in representing all computations. Unfortunately, these models only assure feasibility in computation representation and fail to provide primitive operators to program complex algorithms. 



Keeping performance concerns in mind, the researchers present several computation models and corresponding languages providing operators for specific domains, such as relation algebra~\cite{aho1979universality}, SQL~\cite{sql} for database, and MATLANG~\cite{geerts2021expressive} for scientific computation. Besides them, there are also implicit models and operators provided by the libraries or computation engines like TensorFlow~\cite{abadi2016tensorflow} and PyTorch~\cite{paszke2019pytorch} for deep learning. Their intuition is to define atomic computations, offer language with operators that are compositions of one or more atomic computations, and implement every operator to ensure the best performance on major computation tasks. However, these models or corresponding operators are confronted with two major problems. First, they are designed for a relatively small domain. For example, relation algebra is only suitable for structured data. As a result, they remain low compatibility and high portability cost issues when extending those models to support emerging computing. Second, they have poor generalized expression ability. The generalized expression ability depicts the ability to represent newly-added computations on the existing computation model. If users want to implement new computations in a particular model with poor generalized expression ability, they must implement new operators case by case. As a result, there are more and more newly-defined complicated operators with different semantics. 

To attack the above challenges, this article proposes a unified computation model with the generalized expression ability and a concise set of primitive operators. First, we propose a unified data abstraction -- Tensor of List. Tensor of List is a data abstraction with generalized expression ability, and classical data abstractions like vector, matrix, tensor, and list, can be seen as its particular instances. Based on Tensor of List, we present a unified computation model. The proposed computation model introduces five atomic computations. The five atomic computations can represent any elementary computation on the ToL data abstraction by finite composition, ensured with strict formal proof. We call the Tensor of List data abstraction and the corresponding computation abstraction the ToL model (In short, ToL). In the rest of this article, ToL refers to the ToL data abstraction or the ToL model in different contexts. Based on ToL, we design a  pure-functional language -- ToLang, which provides a concise set of primitive operators that can be used to program complex big data and AI algorithms. 


We focus on four aspects to evaluate how our computation model solves the problems. Table~\ref{tab1} is the overall comparison of different models/languages. As for the applicable domain, ToL suits all domains and provides three levels of generalized expression ability: data abstraction, a limited composition of the five atomic computations that can represent any ToL elementary computation, and a concise set of primitive operators. As for conciseness, we show a smaller set of primitive operators in ToLang can present a large set of ONNX~\cite{onnx} operators in the AI domain. Finally, as for performance concerns, we show ToL has a built-in performance indicator, born with a strictly defined computation metric -- elementary operation count (EOPs), consistent with FLOPs~\cite{2010The} within a small error range.



\begin{table}[ht]
\centering
\resizebox{\linewidth}{!}{
\begin{tabular}{llllccc}
\hline
Model   & Language         & Data abstractions     & Atomic computations (quantity)  & Domain        & \tabincell{c}{Generalized\\expression\\ability} &  \tabincell{c}{Operator\\quantity} \\\hline
Turing machine & -  & Finite array         & Read, write, lshift, rshift (4)  & All                  & high                         & -                \\\hline
Lambda calculus & - & Function             & Lambda (1)                 & All                  & high                          & -                \\\hline
Relation algebra & SQL & Structured data      & \tabincell{l}{Select, project, Cartesian,\\union, intersection, rename} (6)            & Structured         & high                         & 10+                \\\hline
Matlang  & Matlab        & Matrix               & \tabincell{l}{Variable, transpose, one-vector,\\diagonalize, multiply, add, scalar,\\point-wise} (8)           & Linear               & middle                       & 100+              \\\hline
Arithmetic circuits & - & Matrix & Add, sub, mul,... (10+) & Linear & low & - \\\hline
-   & Haskell             & All types            & Add, sub, mul, max, ... (100+)        & All            & low                          & 100+         \\\hline
-   & \tabincell{l}{TensorFlow,\\PyTorch}             & Tensor               & Conv, ReLU, MatMul, ... (100+)        & Tensor       & low                          & 100+            \\\hline
ToL & ToLang       & Tensor of List & Lambda, norm, member, join, embed (5)   & All             & high                                & 13                 \\\hline    
\end{tabular}}
\caption{The comparisons of different models/languages. ToL has a generalized expression ability. It has only five atomic computations and 10+ primitive operators, suiting all application domains.}
\label{tab1}
\end{table}

The organization of this article is as follows. Section 2 is the related work. We make all the necessary definitions for ToL data and computation abstractions in Sections 3.1-3.3. We present strict proof for ToL's completeness and atomicity properties in Section 3.4. We present ToLang in Section 4. We evaluate ToLang's expression ability in Section 5. We demonstrate ToL's built-in performance indicator in Section 6. Section 7 is the conclusion.


%% file: 2.Related.tex
\section{Related work}

The earliest computation models in the computer science domain are Turing machine~\cite{turing1936computable}, Boolean circuits~\cite{vollmer1999introduction}, and lambda calculus~\cite{turing1937computability}. They have equivalent abilities in representing all computations but fail to provide primitive operators for programming complex algorithms. Late in different sub-domains, several data and computation models are proposed without generalized expression abilities to represent newly-added computations. Typical data abstractions include matrix, widely used in linear algebra,  tensor, used in deep learning, set and tuple, in database~\cite{kline1990mathematical}. Typical computation models have relation algebra~\cite{aho1979universality} and tuple relation calculus~\cite{codd1970relational} in the database, tensor computation in the AI, and matrix computation in the scientific computation domains, respectively.

Our proposed ToL model has a generalized expression ability and a concise set of primitive operators. Suiting all application domains, it has only five atomic computations and 10+ primitive operators: a limited composition of the five atomic computations can represent any ToL elementary computation, and a concise set of primitive operators can program complex big data and AI algorithms.  

Besides computation models, the researchers characterized primary computation characteristics in big data, AI, and other domains by workload characterizations. First, the Berkeley multi-disciplinary teams~\cite{asanovic2006landscape} classify computation of different algorithms into 13 basic modes, named data dwarfs. After that, Gao et al.~\cite{gao2018data} investigated five application domains of big data and AI and decomposed various types of computation into eight classes of computation units, named data motifs.

As for high-level language, there are some description languages based on data abstractions and computation models, including SQL language~\cite{melton1993understanding} based on relation algebra, MATLANG~\cite{brijder2017expressive} based on linear matrix computation, and Haskell~\cite{hudak1992gentle} as a pure-functional language for general purpose.

%% file: 3.ToL.tex
\section{ToL}
This section presents ToL, a computation model that can express all elementary computations in different domains. Our methodology is as follows. First, we define the ToL data abstraction. Second, we present the ToL atomic computations: a minimum set of atomic computations defined on ToL. Third, we prove all elementary computations can be expressed using a  finite composition of ToL atomic computations. 


\subsection{ToL Data abstraction}
\textbf{Definition 1.1 (Type space)}: A \textbf{type space} is defined as a set, noted as $V$. An element in a type space is defined as a \textbf{numeric}, noted as $v\in V$. The set of all type spaces is defined as the \textbf{type space set}, noted as $\mathbb{V}=\{V|V$ is a type space$\}$.

\textbf{Definition 1.2 (List)}: A \textbf{list space} is defined as the Cartesian product of a finite quantity of type space, noted as $L=V_1\times V_2\times\cdots\times V_k$, for $V_j\in\mathbb{V}, 1\leq j\leq k, k\in \mathbb{N^*}$. We note $[V_1,V_2,\cdots,V_k]$ as the \textbf{type list} and $k$ as the \textbf{capacity} of list. An element in a list space is defined as a \textbf{list}, noted as $l\in L$. The set of all list spaces is defined as the \textbf{list space set}, noted as $\mathbb{L}=\{L|L$ is a list space$\}$. A list is noted as $(l_1,\cdots,l_k)$.

\textbf{Definition 1.3 (Tensor of List, ToL)}: A \textbf{ToL space} is defined as a multidimensional extension of list spaces, noted as $T=L^{d_1\times d_2\times\cdots\times d_n}$, for $d_i\in\mathbb{N^*}, 1\leq i\leq n, n\in\mathbb{N^*}$. We note $[d_1,d_2,\cdots,d_n]$ as the \textbf{shape} and $n$  the \textbf{dimension} of ToL, respectively. An element in a ToL space is defined as a \textbf{ToL}, noted as $t\in T$. The set of all ToL spaces is defined as the \textbf{ToL space set}, noted as $\mathbb{T}=\{T|T$ is a ToL space$\}$. Numeric, vector, matrix, tensor, and list are ToL instances. The capacity of a tensor is 1. A ToL is noted as $[\cdots[t_{1,\cdots,1,1},...,t_{1,\cdots,1,d_n}],...,t_{1,\cdots,d_{n-1},1},...,t_{1,\cdots,d_{n-1},d_n}],\cdots,t_{d_1,d_2,\cdots,d_n}]$.

\textbf{Example 1}: Real set $\mathbb{R}\in\mathbb{V}$ is a type space.

$L=\mathbb{Z}\times\mathbb{R}\times\mathbb{N^*}$ is a list space with 3 capacity. List $l=(0,3.0,1)\in L$.

Numeric $3.14$ is a trivial list with capacity 1.

$T=L^{2\times2}$ is a ToL space with two dimensions. ToL $t=[[(1,1.0,1),(2,2.0,2)],[(3,3.0,3),(4,4.0,4)]]\in T$.

\subsection{ToL elementary computations}\label{elementary_computation}

\textbf{Definition 2.1 (Basic elementary functions)}: According to Liouville~\cite{liouville1833premier}, a basic elementary function is ``a function of several variables that is defined as taking sums, products, roots, polynomial, rational, trigonometric, hyperbolic, exponential functions, and their inverse functions''. All the basic elementary functions in ToL model are listed in Table~\ref{tab2}.

\textbf{Definition 2.2 (Elementary functions)}: According to Liouville~\cite{liouville1833premier}, an elementary function is ``the sum, product, or composition of a finite quantity of basic elementary functions''.

\begin{table}[ht]
\centering
\begin{tabular}{llc}
\hline
function                                                             & parameters & space   \\\hline
add(+), sub(-)                          & $<$var$>$, $<$var$>$ & $\mathbb{C}$ \\
mul($\times$), div($\div$)              & $<$var$>$, $<$var$>$ & $\mathbb{C}$ \\
mod($\%$)                               & $<$var$>$, $<$var$>$ & $\mathbb{Z}$ \\
pow(\^{}), log                          & $<$var$>$, $<$var$>$ & $\mathbb{R}$ \\
and, or, xor                            & $<$var$>$, $<$var$>$ & $\mathbb{B}$ \\
not                                     & $<$var$>$            & $\mathbb{B}$ \\
lt(\textgreater), gt(\textless), eq(=)  & $<$var$>$, $<$var$>$ & $\mathbb{R}$ \\
le($\leq$), ge($\geq$), ineq($\neq$)    & $<$var$>$, $<$var$>$ & $\mathbb{R}$ \\
max, min                                & $<$var$>$, $<$var$>$ & $\mathbb{R}$ \\
sgn                                     & $<$var$>$            & $\mathbb{R}$ \\
sin, cos, tan, asin, acos, atan         & $<$var$>$            & $\mathbb{R}$ \\
sinh, cosh, tanh, asinh, acosh, atanh   & $<$var$>$            & $\mathbb{R}$\\
rand                                    & $<$var$>$            & $\mathbb{R}$\\\hline
\end{tabular}
\caption{The list of basic elementary functions.}
\label{tab2}
\end{table}

\textbf{Definition 2.3 (ToL elementary computations)}: A ToL elementary computation is a function $F\in(\mathbb{T},\mathbb{T},\cdots,\mathbb{T})\mapsto\mathbb{T}$, noted as $F(t_1,t_2,\cdots,t_p;c_1,c_2,\cdots,c_q)=f_0$, where $t_1,t_2,\cdots,t_p,f_0\in\mathbb{T},c_1,c_2,\cdots,$\\$c_q$ are constants, $p,q\in\mathbb{Z+}$, and numeric values in result $f_0$ can be computed with elementary functions with only the following inputs: part of values from certain $t_i$; capacity, shape, dimension of certain $t_i$; certain constant value $c_j$. In other words, ToL elementary computations are elementary functions computed on ToLs.

\textbf{Example 2}: For real matrices $A$ and $B$ of $[n,n]$ shape, $F(A,B)=A+B$ is an elementary computation on ToL. In this example, $F\in(T,T)\mapsto T, T=\mathbb{R}^{n\times n}$. The basic elementary function involved is add (+).

\subsection{ToL atomic computations}\label{atomic_computation}

In this subsection, we state that five ToL atomic computations are the smallest set enough for expressing all ToL elementary computations. 

We suppose ToL $t=t_1=L^{d_1\times d_2\times\cdots\times d_n}$, $t_2=L^{d_1\times d_2\times\cdots\times d_{n-1}\times D_n}$, where $L=V_1\times V_2\times\cdots\times V_k$ in this subsection.

\textbf{Definition 3.1 (lambda)}: A lambda is noted as $\Lambda(f;e_1,e_2,\cdots,e_p;c_1,c_2,\cdots,c_q)$, where $e_1,e_2,\cdots,e_p$ are numeric, $c_1,c_2,\cdots,c_q$ are constants, $f$ is an elementary function, $p,q\in\mathbb{N^*}$. $\Lambda(f;e_1,e_2,\cdots,e_p;c_1,c_2,$\\$\cdots,c_q)$ computes $f(e_1,e_2,\cdots,e_p;c_1,c_2,\cdots,c_q)$ and returns the result. The function $f$ is a basic elementary function or the  composition of several basic elementary functions in Table~\ref{tab2}.

\textbf{Definition 3.2 (norm)}: A norm is noted as $\Vert t\Vert$ or $\Vert t\Vert'$. $\Vert t\Vert=[d_1,d_2,\cdots,d_n]$ and $\Vert t\Vert'=k$. Additionally, $\Vert(\Vert t\Vert)\Vert=n$. 

\textbf{Definition 3.3 (member)}: A member is noted as $\Pi(t;a_1,a_2,\cdots,a_n)$ or $\Pi'(t;j)$, where $a_1,a_2,\cdots,a_n\in\mathbb{N^*},1\leq a_i\leq d_i,1\leq i\leq n,1\leq j\leq k$, and $n$ is the dimension of $t$. $\Pi$ returns the list member of coordinates given by $a_1,a_2,\cdots,a_n$ and $\Pi'$ returns the corresponding ToL with the same shape as $t$ and type list=$V_j$. $\Pi(t;a_1,a_2,\cdots,a_n)$ is also noted as $t[a_1,a_2,\cdots,a_n]$.

\textbf{Definition 3.4 (join)}: A join is noted as $T_1\bowtie T_2$ or $T_1\overline{\bowtie}T_2$. For $T_1\bowtie T_2$, ToLs $t_1$ and $t_2$ must have the same type list, shape, and dimensions except for the last dimension. $t_1\bowtie t_2$ has the same type list, dimension, and shape but the last dimension=$d_n+D_n$. We have \resizebox{.5\columnwidth}{!}{$(t_1\bowtie t_2)[a_1,a_2,\cdots,a_n]=\begin{cases}t_1[a_1,a_2,\cdots,a_n], a_n\leq d_n\\t_2[a_1,a_2,\cdots,a_{n-1},a_n-d_n+1], a_n>d_n\end{cases}$} for $1\leq a_i\leq d_i, 1\leq i\leq n-1$ and $1\leq a_n\leq d_n+D_n$. For $\overline{\bowtie}$, ToL $t_1$ and $t_2$ must have the same dimension and shape. Similar to $\bowtie$, $T_1\overline{\bowtie}T_2$ joins list members of the same coordinates of $t_1$ and $t_2$.

\textbf{Definition 3.5 (embed)} An embed is noted as $\Xi(t)$. $\Xi(t)$ returns a ToL with one more dimension than $t$, the last dimension $d_{n+1}=1$. The relation between $\Xi(t)$ and $t$ is: $\Xi(t)[a_1,a_2,\cdots,a_n,1]=t[a_1,a_2,\cdots,a_n]$.

\textbf{Example 3}: Lambda: $\Lambda(+,e_1,c_1)$, $\Lambda(\times,e_1,e_2)$, $\Lambda(\arcsin,e_1)$, in which $e_1,e_2$ are variable and $c_1$ is constant.\\
Norm: $\Vert[(1,1.),(2,2.),(3,3.)]\Vert=3$. $\Vert[(1,1.),(2,2.),(3,3.)]\Vert'=2$.\\
Member: $t=[[1,2],[3,4]]$, $t[2,2]=4$.\\
Join: $[1,2,3]\bowtie[1,2]=[1,2,3,1,2]$. $(3,0.14)\overline{\bowtie}3.2=(3,0.14,3.2)$.\\
Embed: $\Xi([1,2,3])=[[1],[2],[3]]$.

\begin{figure*}[ht!]
\centering
\includegraphics[width=1\textwidth]{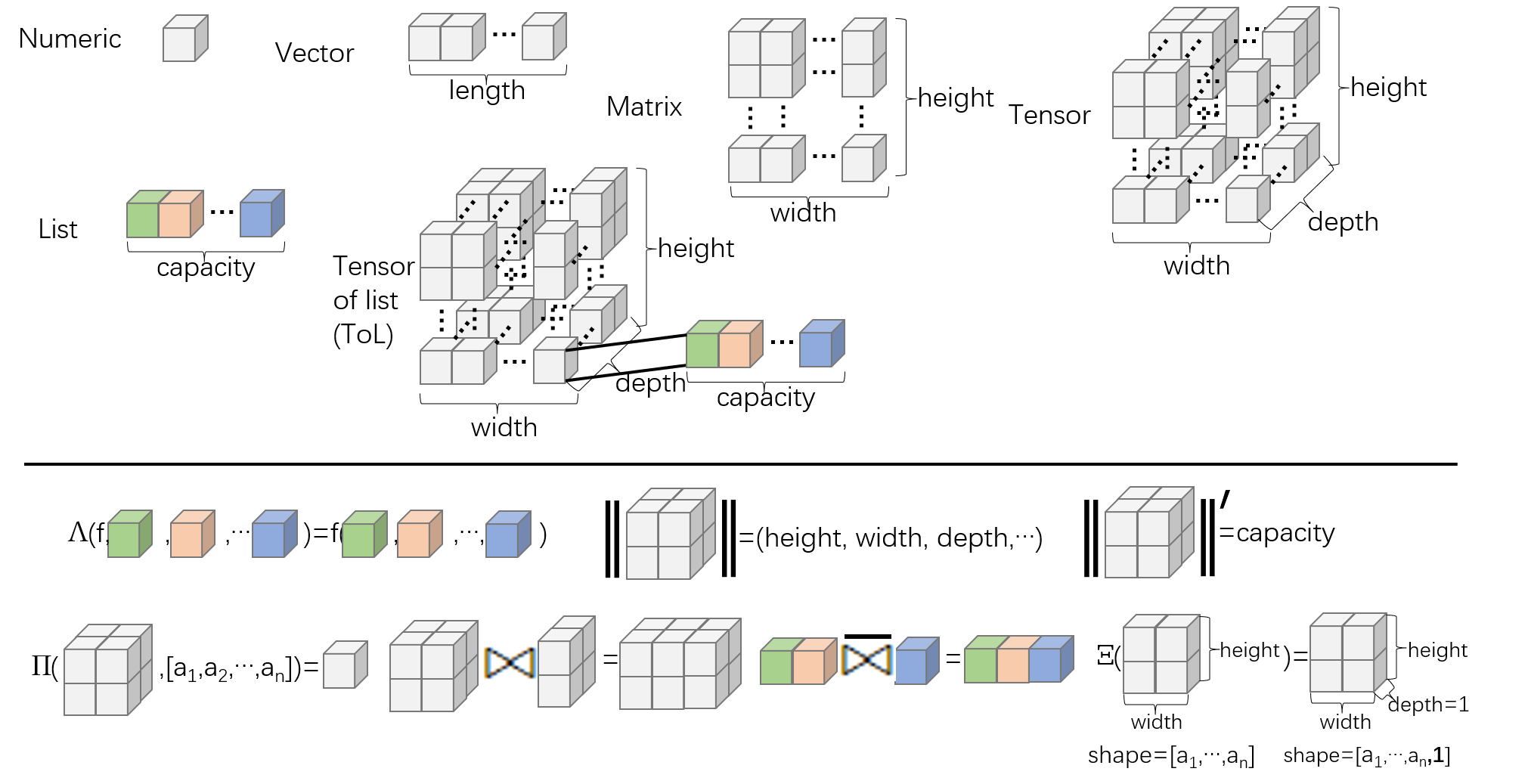}
\caption{The visualizations of ToL, ToL instantiations, and ToL atomic computations.}
\label{fig1}
\end{figure*}

Figure~\ref{fig1} visually shows ToL, ToL instantiations, and ToL atomic computations. ToL instantiations include numeric, vector, matrix, tensor, and list.

\subsection{Proof}\label{section34}
This section proves that the ToL atomic computations  have two properties:  atomicity and completeness.

\textbf{Theorem One: the atomic property}: Each ToL atomic computation cannot be represented by the other atomic computations or their composition.

\textbf{Theorem Two: the complete property}: Any ToL elementary computation can be represented by a finite composition of ToL atomic computations.

The above theorems ensure that five ToL atomic computations are enough for expressing all ToL elementary computations.

\subsubsection{Proof for Theorem One}
We prove that each of the five ToL atomic computations (lambda, norm, member, join, embed) cannot be represented by the other ToL atomic computations or their composition. We construct an example of each ToL atomic computation and show that the rest of the ToL computations cannot represent it.

\textbf{Lambda}: Consider $F(a,b)=a+b$. $\Lambda(+,a,b)$ can represent $F$. However, either norm, member, join, or embed cannot perform addition.

\textbf{Norm}: Consider that $F$ computes the size of vector $v$, $v\in\{[0],[0,0],[0,0,0],\cdots\}$. $F(v)=\Vert v\Vert$. A lambda can compute vector values (zeros), but it cannot return the size of vectors. Member, join, or embed can only return vectors of 0 values and cannot return the size of vectors, either.

\textbf{Member}: Consider that $F$ computes the sum of vector $v$ member values, and $v=[1,2,3,4]$. $F(v)=\Lambda(+,v[1],\Lambda(+,v[2],\Lambda(+,v[3],v[4])$. However, lambda only accepts numeric and function inputs, but $v$ is a vector. Join and embed can only return a ToL, which is not numeric.

\textbf{Join}: Consider $F(t_1,t_2)=t_1\bowtie t_2$, where $t_1$ and $t_2$ are vectors. $\Vert t_1\bowtie t_2\Vert[1]=\Vert t_1\Vert[1]+\Vert t_2\Vert[1]$. The first dimensions of the shapes are added up. However, lambda, norm, and member return numeric values. Embed cannot change the first dimension of a vector.

\textbf{Embed}: Consider $F(t)=\Xi(t)$, where $t$ is a $n$-dimensional tensor. The dimension of $F(t)$ is bigger than that of $t$. However, lambda, norm, and member return numeric values. Join only changes the last dimension and cannot increase the dimension.

\subsubsection{Proof for Theorem Two}
Consider a ToL elementary computation noted as $F(t_1,t_2,\cdots,t_p;c_1,c_2,\cdots,c_q)=f_0$. We discuss by categorization:

\textbf{Case 1}: $f_0$ is a numeric value.

\textbf{Sub-case 1.1}: $t_1,t_2,\cdots,t_p$ are numeric values, and Function $F$'s inputs are them and constants. In this sub-case, we represent an elementary function $F$ with a finite composition of basic elementary functions. Note $F(t_1,t_2,\cdots,t_p;c_1,c_2,\cdots,c_q)=f_1(f_2(\cdots f_r(\cdots,y_r),\cdots,y_2),\cdots,y_1)$, where $y_1,y_2,\cdots,y_r,\cdots$ are rearranged $t_i$'s and $c_j$'s, $r\in\mathbb{N^*}$, and $f_1,f_2,\cdots,f_r$ are basic elementary functions. Then computation $f_0=F(t_1,t_2,\cdots,t_p;c_1,c_2,\cdots,c_q)$ can be represented as $f_0=\Lambda(f_1,\Lambda(f_2,\cdots\Lambda(f_r,y_r,\cdots),y_2,\cdots),y_1,\cdots)$.

\textbf{Sub-case 1.2}: Some of $t_1,t_2,\cdots,t_p$ are not pure numeric but normal ToLs, and Function $F$'s inputs are numeric elements from certain ToLs and constants. Note the numeric inputs of Function $F$ as $x_j$'s, $s\in\mathbb{N^*}$. Then it can be represented as $F(x_1,x_2,\cdots,x_s;c_1,c_2,\cdots,c_q)$. For each $x_j$, if it derives from a certain ToL $t_i$, use $x_j=\Pi(t_i;a_{j_1},a_{j_2},\cdots,a_{j_n})$ for ToL elements and $x_j=\Pi'(t_i;j_m)$ for list elements, where $a_{j_1},a_{j_2},\cdots,a_{j_n},m,n\in\mathbb{N^*}$. Then this sub-case turns to sub-case 1.1.

\textbf{Sub-case 1.3}: Function $F$' inputs include not only numeric variables, ToL's numeric elements, and constants but also the capacity, shape, and dimension of certain ToLs. Suppose $F$'s parameters include capacity, shape, and dimension of a certain $t_i$. We use the following representation:\\
capacity($t_i$)=$\Vert t_i\Vert'$\\
shape($t_i$)=$\Vert t_i\Vert$\\
dimension($t_i$)=$\Vert\Vert t_i\Vert\Vert$\\
Then this sub-case turns to sub-case 1.1 or 1.2.

\textbf{Case 2}: $f_0$ is not a numeric value but a normal ToL.

Note $f_0=T=L^{d_1\times d_2\times\cdots\times d_n}$, $d_i\in\mathbb{N^*}, 1\leq i\leq n, n\in\mathbb{N^*}$ and $L=V_1\times V_2\times\cdots\times V_k, V_j\in\mathbb{V}, 1\leq j\leq k, k\in \mathbb{N^*}$. We construct $f_0$ by the following steps.

\textbf{Step 1}: First we consider a numeric member of $f_0$ at certain coordinates, such as $\Pi'(f_0[a_1,a_2,\cdots,a_n],j)$. According to case 1, it can be represented with lambda, norm, and member computation; we note such representation as $\Omega(a_1,a_2,\cdots,a_n;j)$ in short.

\textbf{Step 2}: For all $j$ that $1\leq j\leq k$, construct list member $f_0[a_1,a_2,\cdots,a_n]=\Omega(a_1,a_2,\cdots,a_n;1)\overline{\bowtie}\Omega(a_1,a_2,$\\$\cdots,a_n;2)\overline{\bowtie}\cdots\overline{\bowtie}\Omega(a_1,a_2,\cdots,a_n;k)$. So far every $f_0[a_1,a_2,\cdots,a_n]$ is represented.

\textbf{Step 3}: For the last dimension $d_n$, construct $f_0[a_1,a_2,\cdots,a_{n-1}]=f_0[a_1,a_2,\cdots,1]\bowtie f_0[a_1,a_2,\cdots,2]\bowtie\cdots\bowtie f_0[a_1,a_2,\cdots,d_n]$. Therefore, $\Vert f_0[a_1,a_2,\cdots,a_{n-1}]\Vert=d_n$. If $n=1$, the representation is complete.

\textbf{Step 4}: If $n>1$, construct $\Xi(f_0[a_1,a_2,\cdots,a_{n-1}])$, noted as $f_0'[a_1,a_2,\cdots,a_{n'}]$ for short, where $n'=n-1$, and $\Vert f_0'[a_1,a_2,\cdots,a_{n'}]\Vert=[d_n,1]$. Continue to construct $f_0'[a_1,a_2,\cdots,a_{{n'}-1}]=f_0'[a_1,a_2,\cdots,1]\bowtie f_0'[a_1,a_2,\cdots,2]\bowtie\cdots\bowtie f_0'[a_1,a_2,\cdots,d_{n'}]$. Repeat steps 3 and 4 until all the dimensions of $f_0$ are represented.

\textbf{Example 4}: Computation $F(a,b)=f_0$ computes matrix multiplication of matrices $a$ and $b$ sized $[2,2]$. It can be represented with atomic computations as follows:\\
$f_{1,1}=\Lambda(+,\Lambda(\times,a[1,1],b[1,1]),\Lambda(\times,a[1,2],b[2,1]))$\\
$f_{1,2}=\Lambda(+,\Lambda(\times,a[1,1],b[1,2]),\Lambda(\times,a[1,2],b[2,2]))$\\
$f_{2,1}=\Lambda(+,\Lambda(\times,a[2,1],b[1,1]),\Lambda(\times,a[2,2],b[2,1]))$\\
$f_{2,2}=\Lambda(+,\Lambda(\times,a[2,1],b[1,2]),\Lambda(\times,a[2,2],b[2,2]))$\\
$f_0=\Xi(f_{1,1}\bowtie f_{1,2})\bowtie\Xi(f_{2,1}\bowtie f_{2,2})$

%% file: 4.ToLang.tex
\section{ToLang}\label{Section:ToLang}
This section proposes a ToL-based language -- ToLang. ToLang is a pure-functional language and can proceed with any ToL elementary computation. The statements and primitive operators are based on the compositions of five ToL atomic computations. 

\subsection{Data declaration}\label{section41}
ToLang supports a type space declaration in the following formats:\\
\texttt{\textbf{space} <var>: <space>}\\
\texttt{\textbf{space} <var>: <space> [<lower\_bound>, <upper\_bound>]}\\
In the two formats, \texttt{\textbf{space}} is a reserved word for declaration, \texttt{<var>} is a variable name, \texttt{<space>} is a defined type space name, \texttt{<lower\\\_bound>} and \texttt{<upper\_bound>} are bounds of the data range. The predefined type spaces are the following:\\
\texttt{B N N* Z Z* Z+ Z- Q Q* Q+ Q- R R* R+ R- I C}\\
which stands for:\\
$\mathbb{B}$, $\mathbb{N}$, $\mathbb{N^*}$, $\mathbb{Z}$, $\mathbb{Z^*}$, $\mathbb{Z^+}$, $\mathbb{Z^-}$, $\mathbb{Q}$, $\mathbb{Q^*}$, $\mathbb{Q^+}$, $\mathbb{Q^-}$, $\mathbb{R}$, $\mathbb{R^*}$, $\mathbb{R^+}$, $\mathbb{R^-}$, $\mathbb{I}$, $\mathbb{C}$.

ToLang supports a list declaration in the following formats:\\
\texttt{\textbf{list} <var>: <list>}\\
\texttt{\textbf{list} <var>: <space>}\\
\texttt{\textbf{list} <var>: [<space>, <space>, $\cdots$], <space>}\\
In the three formats, \texttt{\textbf{list}} is a reserved word for declaration, and \texttt{<list>} is a defined list name. In the third format, the quantity of \texttt{<space>}'s can be any finite positive integer.

ToLang supports a Tensor of List declaration. ToL is the most critical data structure, whose instantiations include numeric, vector, matrix, tensor, and list. In ToLang, any computation is defined and computed on ToLs. They are in the following formats:\\
\texttt{\textbf{tol} <var>: <space>} // for numeric\\
\texttt{\textbf{tol} <var>: <space> [<length>]} // for vector\\
\texttt{\textbf{tol} <var>: <space> [<height>, <width>]} // for matrix\\
\texttt{\textbf{tol} <var>: <space> [<length>, <length>, $\cdots$, <length>]} // for tensor\\
\texttt{\textbf{tol} <var>: <list>} // for list\\
\texttt{\textbf{tol} <var>: <list> [<length>, <length>, $\cdots$, <length>]} // for any type of ToL\\
In the above formats, \texttt{<length>}, \texttt{<height>}, and \texttt{<width>} must be a positive integer.

\subsection{ToL primitive operators}\label{section42}

ToLang includes several predefined ToL primitive operators that are the compositions of one or more ToL atomic computations~\ref{atomic_computation}. They are defined on ToLs and require only (one or several) ToLs and functions as their inputs. The mappings between ToL primitive operators and ToL atomic computations are shown in Table~\ref{tab4}. We present each ToL operator as follows. In the rest of this article, when we refer to a ToL atomic computation, we only mention its name, e.g., lambda.

\begin{table}[ht]
\centering
\resizebox{\linewidth}{!}{
\begin{tabular}{lll}
\hline
ToL primitive operator & ToL atomic computations & Mapping\\\hline
function     & lambda                       & \texttt{f(v\_1,v\_2,$\cdots$,c\_1,c\_2,$\cdots$)}=$\Lambda$\texttt{(f;v\_1,v\_2,$\cdots$,c\_1,c\_2,$\cdots$)}\\\hline
shape        & norm                         & \texttt{shape((t\_i))}=\texttt{$\Vert$t\_i$\Vert$}\\\hline
dimension    & norm                         & \texttt{dimension(t\_i)}=$\Vert\Vert$t\_i$\Vert\Vert$\\\hline
capacity     & norm                         & \texttt{capacity(t\_i)}=$\Vert$t\_i$\Vert'$\\\hline
volume       & norm                         & \texttt{vol($t$)}=$\Vert t\Vert[1]\times\Vert t\Vert[2]\times\cdots\times\Vert t\Vert[$\texttt{dim(}$t$\texttt{)}$]\times\Vert t\Vert'$\\\hline
space        & member                       & space($t$)$\mapsto\mathbb{V}$\\\hline
convert      & member, join, embed          & convert($t$,$\mathbb{V'}$)$\mapsto t$ in $\mathbb{V'}$\\\hline
part         & member, join, embed          & \tabincell{l}{\texttt{part($t$,[$a_1,a_2,\cdots,a_n$], $b_1,b_2,\cdots,b_n$])}=$\Xi(\cdots\Xi(\Pi(t;a_1,a_2,\cdots,a_n)$\\$\bowtie\Pi(t;a_1+1,a_2,\cdots,a_n)\bowtie\Pi(t;a_1+2,a_2,\cdots,a_n)\bowtie\cdots\Pi(t;b_1,a_2,\cdots,$\\$a_n)\bowtie\Pi(t;b_1,a_2+1,\cdots,a_n)\bowtie\cdots\Pi(t;b_1,b_2,\cdots,b_n)\cdots)$}\\\hline
swap         & member, join, embed          & \tabincell{l}{\texttt{swap($t$,dim,[$x_1,x_2,\cdots,x_k$])}=$\Xi(\cdots\Xi(\Pi(t;1,1,\cdots,x_1,\cdots,1)\bowtie\Pi(t;1,1,$\\$\cdots,x_2,\cdots,1)\bowtie\cdots\Pi(t;1,1,\cdots,x_k,\cdots,1)\bowtie\cdots\Pi(t;a_1,a_2,\cdots,x_1,\cdots,$\\$a_n)\bowtie\Pi(t;a_1,a_2,\cdots,x_2,\cdots,a_n)\bowtie\cdots\Pi(t;a_1,a_2,\cdots,x_k,\cdots,a_n)\cdots)$}\\\hline
reshape      & member, join, embed          & \tabincell{l}{\texttt{reshape($t$,shape)}=$\Xi(\cdots\Xi(\Pi(t;1,1,\cdots,1)\bowtie\Pi(t;1,1,\cdots,2)\bowtie\cdots$\\$\Pi(t;1,1,\cdots,a_1)\bowtie\cdots\Pi(t;a_1,a_2,\cdots,a_n)\cdots)$}\\\hline
tile         & member, join, embed          & \tabincell{l}{\texttt{tile($t$)}=$\Xi(\cdots\Xi(\Pi'(\Pi(t;1,1,\cdots,1),1)\bowtie\Pi'(\Pi(t;1,1,\cdots,1),2)\bowtie\cdots$\\$\Pi'(\Pi(t;1,1,\cdots,1),k)\bowtie\cdots\Pi'(\Pi(t;a_1,a_2,\cdots,a_n),k)\cdots)$}\\\hline
map          & lambda, member, join, embed  & \tabincell{l}{\texttt{map($t$,$f$,iter)}=$\Xi(\cdots\Xi(\Lambda(f,t[1,1,\cdots,1]))\bowtie\Xi(\Lambda(f,t[2,1,\cdots,1]))$\\$\bowtie\cdots\bowtie\Xi(\Lambda(f,t[d_1,1,\cdots,1])))\bowtie\cdots\bowtie\Xi(\Lambda(f,t[d_1,d_2,\cdots,d_n])\cdots)$}\\\hline
reduce       & lambda, member, join, embed  & \tabincell{l}{\texttt{reduce($t$,$f$,$i$,iter)}=$\Lambda(f,t[d_1,d_2,\cdots,d_n],\cdots\Lambda(f,t[d_1,1,\cdots,1],\cdots$\\$\Lambda(f,t[2,1,\cdots,1],\Lambda(f,t[1,1,\cdots,1],i)\cdots)$}\\\hline
\end{tabular}}
\caption{This table lists the mappings between  each ToL primitive operator and the composition of one or more ToL atomic computations.}
\label{tab4}
\end{table}

\textbf{Function}: a function \texttt{f()} with several input variables \texttt{v\_1, v\_2, $\cdots$} and constants \texttt{c\_1, c\_2, $\cdots$} is represented in the following formats:\\
\texttt{f(v\_1,v\_2,$\cdots$,c\_1,c\_2,$\cdots$)}\\
\texttt{f.g.h....(v)} //which stands for \texttt{...(h(g(f(v)...)}, composite function\\
The mapping between a function and lambda is as follows:\\
\texttt{f(v\_1,v\_2,$\cdots$,c\_1,c\_2,$\cdots$)}=$\Lambda$\texttt{(f;v\_1,v\_2,$\cdots$,c\_1,c\_2,$\cdots$)}

\textbf{Shape, dimension, capacity}: they are the same as Definitions 1.2 and  1.3 and represented in the following formats:\\
Shape: \texttt{shape(<var>)} or \texttt{|<var>|}\\
Dimension: \texttt{dim(<var>)} or \texttt{||<var>||}\\
Capacity: \texttt{capacity(<var>)} or \texttt{|<var>|'}\\
The mappings between shape, dimension, or capacity operators and norm are as follows:\\
\texttt{shape((t\_i))}=\texttt{$\Vert$t\_i$\Vert$}\\
\texttt{dimension(t\_i)}=$\Vert\Vert$t\_i$\Vert\Vert$\\
\texttt{capacity(t\_i)}=$\Vert$t\_i$\Vert'$

\textbf{Volume}: it is defined as the total quantity of numeric in a certain ToL, represented in the following format:\\
\texttt{vol(<var>)}\\
According to definition, we have \texttt{vol($t$)}=$\Vert t\Vert[1]\times\Vert t\Vert[2]\times\cdots\times\Vert t\Vert[$\texttt{dim(}$t$\texttt{)}$]\times\Vert t\Vert'$.

\textbf{Type list}: it returns the (default) type list of a ToL, represented in the following format:\\
\texttt{space(<var>)}

\textbf{Example 5}: \texttt{space([0,3.0,1])}=\texttt{[Z,R,N]}. Although it may be \texttt{[Z,R*,N]}, \texttt{[Z,R,N*]}, or something else, this operator always returns the default type list.

\textbf{Convert}: it changes the type list of a ToL to a given one, represented in the following format:\\
\texttt{convert(<var>, <type\_list>)}

\textbf{Example 6}: A tensor \texttt{t} is an integer matrix, and \texttt{convert(t,[R])} returns a floating-point matrix with the same values as \texttt{t}.

\textbf{Part}: it returns a sub-ToL of a ToL, represented in the following formats. A part operator is like the slice or index operator in other computer languages.\\
A part operator is a composition of member, join, and embed:\\
// equivalent to $\Pi(t;a_1,a_2,\cdots,a_n)$\\
\texttt{part(<var>,[a\_1,a\_2,$\cdots$,a\_n])}\\
or \texttt{<var>[a\_1,a\_2,$\cdots$,a\_n]}\\
// taking a sub-ToL, equivalent to $\Pi(t;a_1,a_2,\cdots,a_n)\bowtie\Pi(t;a_1+1,a_2,\cdots,a_n)\bowtie\Pi(t;a_1+2,a_2,\cdots,a_n)\bowtie\cdots\Pi(t;b_1,a_2,\cdots,a_n)\bowtie\Pi(t;b_1,a_2+1,\cdots,a_n)\bowtie\cdots\Pi(t;b_1,b_2,\cdots,b_n)$\\
\texttt{part(<var>,[a\_1,a\_2,$\cdots$,a\_n],[b\_1,b\_2,$\cdots$,b\_n])}\\
// equivalent to $\Pi'(t;j)$\\
\texttt{part'(<var>,j)} or \texttt{<var>[j]'}

\textbf{Example 7}: \texttt{part([[1,2],[3,4]],[2,2])=4}.

\textbf{Swap}: it changes ToL's elements arrangement of given coordinates, represented in the following format:\\
\texttt{swap(<var>,<dim>,[x\_1,x\_2,$\cdots$,x\_k])}\\
In the representation, \texttt{<dim>} is an integer indicating the given dimension, \texttt{x\_1,x\_2,$\cdots$,x\_k} are integers from \texttt{1} to \texttt{k} indicating arrangement order, \texttt{k} is the length of a ToL \texttt{<var>}'s \texttt{<dim>}-th dimension. A swap operator is combined with member and join.

\textbf{Example 8}: \texttt{swap([[1,2],[3,4]],1,[2,1])=[[2,1],[4,3]]}.

\textbf{Reshape, Tile}: the two operators change a ToL's shape and type list but do not change its numeric quantity. They are represented in the following formats:\\
// only changing the shape\\
\texttt{reshape(<var>,<shape>)}\\
// tiling the last dimension of ToL to list space\\
\texttt{tile(<var>)}\\
Reshape is easy to understand, for \texttt{<shape>} stands for a new shape of the same data quantity. We give some additional explanation for the tile operator. The ToL tile operator converts its last dimension to its list spaces. We suppose ToL $t$, for example. If $\Vert t\Vert=[4,3,2]$ and the type list of $t$ is $[\mathbb{Z,Q}]$, then $\Vert $\texttt{tile(}$t$\texttt{)}$\Vert=[4,3]$ and the type list of $\texttt{tile(}$t$\texttt{)}$ is $[\mathbb{Z,Q,Z,Q}]$. In other words, a tile operator reforms the last dimension of a ToL into its list space. Reshape and tile operators can be combined with member and join.

\textbf{Example 9}: \texttt{reshape([[1,2],[3,4],[5,6]],[2,3])=[[1,2,}\texttt{3],[4,5,6]]}, \texttt{tile([[(1,'a'),\\(2,'b')],[(3,'c'),(4,'d')]]}\texttt{=[(1,'a',2,'b'),(3,'c',4,'d')]}.

\textbf{Map}: Map is defined as the same function computed on every element of a ToL. It is represented in the following two formats:\\
\texttt{map(<var>,<func>,<iter>)}\\
\texttt{map(<var>,<func\_co>,<iter>)}\\
In the first representation, \texttt{<func>} is a numeric function that requires the elements of ToL \texttt{<var>} as its only parameter. In the second representation, \texttt{<func\_co>} is a numeric function that requires the elements and their coordinates (represented as \texttt{*<var>} or \texttt{<iter>}, not necessary) of ToL \texttt{<var>} as its two parameters.\\
A map operator is combined with lambda, member, join and embed. We present further explanation. For ToL $t$ shaped $[d_1,d_2,\cdots,d_n]$, function $f$, \texttt{map($t$,$f$)} is equivalent to\\
$\Xi(\cdots\Xi(\Lambda(f,t[1,1,\cdots,1]))\bowtie\Xi(\Lambda(f,t[2,1,\cdots,1]))\bowtie\cdots\bowtie\Xi(\Lambda(f,t[d_1,1,\cdots,1])))\bowtie\cdots\bowtie\Xi(\Lambda(f,t[d_1,$\\$d_2,\cdots,d_n])\cdots)$.

\textbf{Reduce}: The reduce operator is defined as the same function computed on every element of a ToL as its inputs. It is represented in the following two formats:\\
\texttt{reduce(<var>,<func>,<var\_0>,<iter>)}\\
\texttt{reduce(<var>,<func\_co>,<var\_0>,<iter>)}\\
In the first representation, \texttt{<func>} is a numeric function with two parameters. Function \texttt{<func>} is firstly computed with initial value \texttt{<var\_0>} and the first element of \texttt{var}, then computed with the last result and the second element of \texttt{var}, and so on, until all elements are traversed. In the second representation, \texttt{<func\_co>} is a numeric function with three parameters. The third parameter is for the coordinates (represented as \texttt{*<var>} or \texttt{<iter>}, not necessary) of every element.\\
The reduce operator is constructed with lambda, member, join, and embed. We present further explanation. For ToL $t$ shaped $[d_1,d_2,\cdots,d_n]$, function $f$, and initial value $i$, \texttt{reduce($t$,$f$,$i$)} is equivalent to $\Lambda(f,t[d_1,d_2,\cdots,d_n],\cdots\Lambda(f,t[d_1,1,\cdots,1],\cdots\Lambda(f,t[2,1,\cdots,1],\Lambda(f,t[1,1,\cdots,1],i)\cdots)$.

\textbf{Example 10}: \texttt{map([1.,2.,3.],sqrt)=[1.0,1.41,1.73]},\\ \texttt{reduce([1,2,3,4,5],add,0)=15}.

\textbf{Example 11}: Suppose $m$ is a matrix, $f_1(m)$ computes the element sum of $m$, and $f_2(m,r)$ computes $m$'s scalar multiplication with real number $r$. Then, $f_1(m)=$reduce($m$,+,0) and $f_2(m,r)=$map($m$,$f_r$), where $f_r(x)=x\times r$. In this example, variable \texttt{<iter>} is left out.




%% file: 5.Expression.tex
\section{Evaluating ToLang's expression ability}

This section evaluates ToL's expression ability from qualitative analysis and practice study. We also provide three case studies to show how to represent widely-used algorithms in scientific computing and deep learning in ToLang. Our evaluations show ToLang has a generalized expression ability.

\subsection{Qualitative Analysis}

To evaluate ToLang's expression ability, we use arithmetic circuits~\cite{shpilka2010arithmetic} for comparison. Arithmetic circuits have no lower expression ability and counting complexity than Boolean circuits. Therefore, it has been used as a scale for expression ability. From three perspectives, we argue that ToL has stronger --at least equivalent -- expression ability than arithmetic circuits.

First, arithmetic circuits are used in the linear algebra domain, whose data abstraction is matrices. However, ToL can be instantiated as more data abstraction than matrices. Second, ToL covers arithmetic computation and can represent basic algorithms in arithmetic circuits. This is proved in Section~\ref{section34}. Third, the composition of ToL operators is defined as finite pure function nesting, equivalent to directed acyclic graph (DAG) and arithmetic circuits.

\subsection{Practice Studies}
ONNX (Open Neural Network Exchange)~\cite{onnx} is an open format built to represent machine learning models. It includes the definition of most classical computations on tensors and other data formats, covering most computations in big data and AI domains. It also includes most linear operators. 
Hereby, we compare the ToL operators against the ONNX operators to demonstrate the former's conciseness. Table~\ref{tab31} and \ref{tab32} listed primary operators with their parameter formats in ONNX. We present their representation by composing ToL primitive operators in the second column in Table~\ref{tab31} and \ref{tab32}. From this table, we can see that 60 primary complicated  ONNX operators (ignoring the other trivial ONNX operators) can be composed with only 13 ToL primitive operators, which shows the conciseness of ToLang.

The existing engines and libraries, such as TensorFlow~\cite{abadi2016tensorflow} and PyTorch~\cite{paszke2019pytorch}, defines various operators for use. However, users must implement new-added operators case by case when supporting new AI models. As a result, the existing engines and libraries may contain newly-defined complicated operators with different semantics for different usage. ToLang, however, does not need to add new operators. As is proved in Section~\ref{section34}, ToL's completeness property ensures that new computations can also be represented with the primitive operators rather than defining new operators case by case.

We present a case study to compare ToLang against TensorFlow on newly-added operators. A user wants to compute the shortest paths of a matrix-stored weight graph $(E_{i,j})_{i,j}$ with Dijkstra's algorithm~\cite{cormen2022introduction}. The following formula represents this algorithm:\\
$E_{i,j}=$\textbf{min}$_k(E_{i,k}+E_{k,j})$\\
In TensorFlow~\cite{abadi2016tensorflow}, the user needs a user-defined function (UDF) for implementation. In ToLang, this algorithm can be represented as follows (name adjacency graph \texttt{E}):\\
\texttt{\textbf{tol} result=[|E|[1],|E|[2]]}\\
\texttt{vec(i,j)=map(E[i],add(*,b[*,j]))}\\
\texttt{elem(i,j)=reduce(vec(i,j),min,+Inf)}\\
\texttt{result:=map(result,elem(i,j),[i,j])}

The above case studies show ToLang has a generalized expression ability in newly emerging scenarios.

\subsection{Case studies: representing important algorithms in ToLang}

We provide three case studies demonstrating how to present classical computations in ToLang. They are matrix multiplication, convolution, and pooling, widely used in deep learning and scientific computation. In addition, we provide the ToLang representations of two complete algorithms in Section~\ref{section54}.

\subsubsection{Matrix multiplication}
\label{section521}
We select the standard definition of matrix multiplication as the baseline. This computation is represented as \texttt{result}=MatMul\texttt{(a,b)} and defined the same as that in numpy.matmul~\cite{oliphant2006guide}. We suppose input matrices \texttt{a} and \texttt{b} are 2-dimensional (matrix), and their size fits. We represent it in ToLang as follows:\\
\texttt{\textbf{tol} result=[|a|[1],|b|[2]]}\\
\texttt{vec(i,j)=map(a[i],mul(*,b[*,j]))}\\
\texttt{elem(i,j)=reduce(vec(i,j),add,0)}\\
\texttt{result:=map(result,elem(i,j),[i,j])}\\
We provide further explanation. Function \texttt{mul(*,y)} is the multiplication function with only the second parameter \texttt{y}. The first line is the definition of \texttt{result} variable, restricting its shape. The second line computes \texttt{vec(i,j)}, which is the multiplication result of elements on the \texttt{i}-th row of matrix \texttt{a} and the \texttt{j}-th column of matrix \texttt{b}. The third line computes the element sum of \texttt{vec(i,j)}, which equals the element of \texttt{result[i,j]}. Finally, the last line computes the result. Figure~\ref{fig3a} is a visualization of ToLang's representation.

\subsubsection{Convolution}
We consider 2-dimensional convolution with a single channel and select the definition of Conv2d in torch.nn~\cite{paszke2019pytorch}. We make some necessary simplifications because the primitive definition considers some implementation factors. The ToLang's representation of this algorithm is as follows: \texttt{result}=Conv2d\texttt{(in,k,\\s,p)}, where \texttt{in,k,s,p} is input, kernel, stride, and padding, respectively. We represent it in ToLang as follows:\\
\texttt{\textbf{tol} result=[(|in|[1]+2*p-|k|[1])/s+1,(|in|[2]+2*p-|k|[2])/s+1]}\\
\texttt{in\_pad=[\textbf{tol}[p,2*p+|in|],[\textbf{tol}[p,|in|],in,\textbf{tol}[p,|in|]],\textbf{tol}[p,2*p+|in|]]}\\
\texttt{in0(i,j)=part(in\_pad,[(i-1)*s+1,(j-1)*s+1],[i*s,j*s])}\\
\texttt{out0(i,j)=map(in0(i,j),mul(*,k(i,j)),[i,j])}\\
\texttt{elem(i,j)=reduce(out0(i,j),add,0)}\\
\texttt{result:=map(result,elem(i,j),[i,j])}\\
We provide further explanation. \texttt{\textbf{tol}[x,y]} is a constant matrix with a shape \texttt{[x,y]}, as is defined in Section~\ref{section41}. The first line is the definition of \texttt{result} variable, determining its shape. The second line is padding, and \texttt{in\_pad} is the padded form of matrix \texttt{in}. The third line is selecting a block (\texttt{in0(i,j)}) of the input matrix for convolution. The 4th line is multiplication with block and kernel, where \texttt{out0} is the result. The 5th line computes the element sum of \texttt{out0(i,j)} as the element of \texttt{result[i,j]}. The last line computes the result. Figure~\ref{fig3b} is a visualization of ToLang's representation.

\subsubsection{Pooling}
\begin{figure*}
	\centering
	\subfloat[Matrix multiplication]{\label{fig3a}    \includegraphics[width=0.8\textwidth]{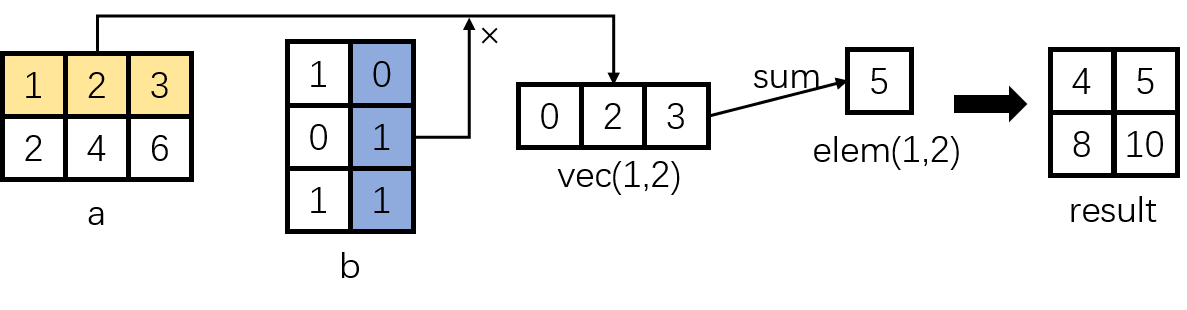}}\\
	\subfloat[Convolution]{\label{fig3b}
	\includegraphics[width=1\textwidth]{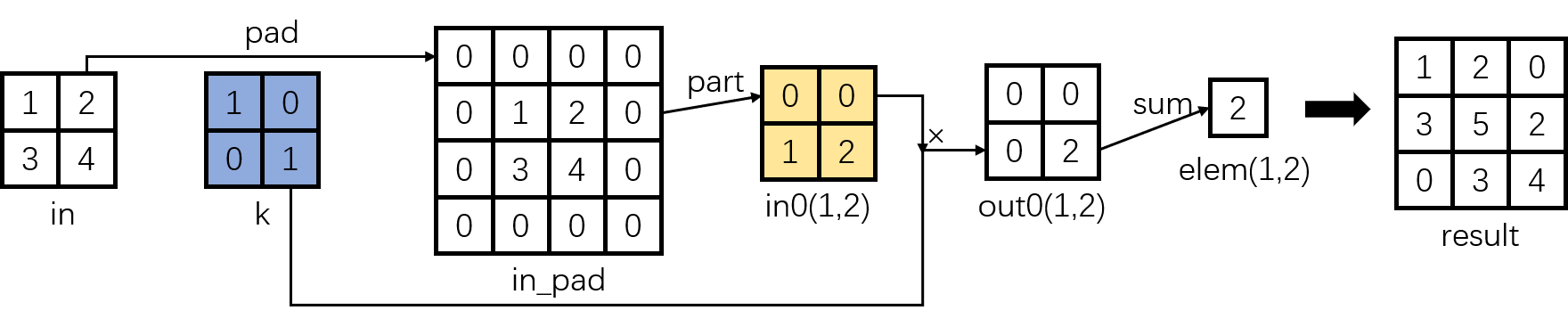}}\\
	\subfloat[Pooling]{\label{fig3c}
	\includegraphics[width=1\textwidth]{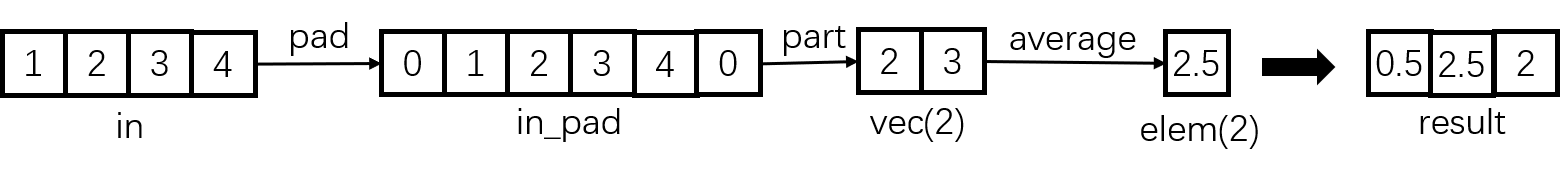}}
	\caption{Visualizations of ToLang's representation of three algorithms.}
	\label{optimization}
\end{figure*}

We consider vector average pooling with a single channel and select the definition of AvgPool1d in torch.nn~\cite{paszke2019pytorch}. We make some necessary simplifications ignoring the implementation details. The ToL's representation of this algorithm is as \texttt{result}=AvgPool1d\texttt{(in,k,s,p)}, where \texttt{in,k,s,p} is input, kernel size, stride, and padding, respectively. We represent it in ToLang as follows:\\
\texttt{\textbf{tol} result=[(|in|[1]+2*p-k)/s+1]}\\
\texttt{in\_pad=[\textbf{tol}[p,|in|],in,\textbf{tol}[p,|in|]]}\\
\texttt{vec(i)=part(in\_pad,[(i-1)*s+1,i*s])}\\
\texttt{elem(i)=reduce(vec(i),add,0)/s}\\
\texttt{result:=map(result,elem(i),i)}\\
We provide further explanation. The first line is the definition of \texttt{result} variable, determining its shape. The second line is padding, and \texttt{in\_pad} is the padded form of vector \texttt{in}. The third line is selecting a block (\texttt{vec(i)}) of the input vector. The 4th line is the computation of its average as \texttt{elem(i)}. Finally, the last line computes the result. Figure~\ref{fig3c} is a visualization of ToLang's representation.

\subsection{Representing algorithms with ToLang}\label{section54}


We present two classical algorithms in ToLang -- Resnet~\cite{he2016deep} and clustering~\cite{krishna1999genetic}. For comparison, also we show the original Python code of each algorithm.



\subsubsection{Representation for Resnet}
\begin{figure}[ht!]
\centering
\includegraphics[width=0.2\textwidth]{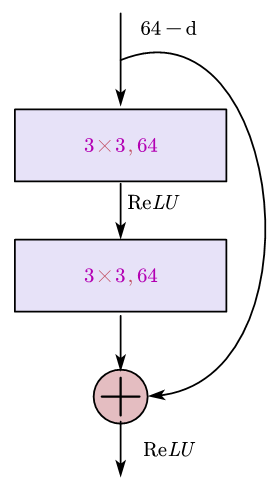}
\caption{The basic block in Resnet model.}
\label{fig4}
\end{figure}

We select a basic block in Resnet model~\cite{he2016deep} as the example. The source code is written in Python with the torch.nn module. Please note that unnecessary details of low-level implementation are removed. Figure~\ref{fig4} is the structure of the basic block from the original paper.

\textbf{Algorithm 1.1: The basic block in Resnet in Python}
\begin{lstlisting}[language=Python]
class BasicBlock(nn.Module):
  def __init__:
   super(BasicBlock, self).__init__()
   self.conv1 = conv3x3(in_channel, kernel_size=3, stride=stride,\
padding=1, bias=False)
   self.bn1 = nn.BatchNorm2d(out_channel)
   self.relu = nn.ReLU(inplace=True)
   self.conv2 = conv3x3(out_channel,kernel_size=3, stride=stride,\
padding=1, bias=False)
   self.bn2 = nn.BatchNorm2d(out_channel)
   self.downsample = downsample
   self.stride =stride
  #block
  def forward(self, x):
   residual = x
   out = self.conv1(x)
   out = self.bn1(out)
   out = self.relu(out)
   out = self.conv2(out)
   out = self.bn2(out)
   if self.downsample is not None:
   residual = self.downsample(x)
   out = out + residual
   out = self.relu(out)
   return out
\end{lstlisting}

\textbf{Algorithm 1.2: The basic block in Resnet in ToLang}\\
\texttt{\#conv}\\
\texttt{\textbf{tol} in=[size,size]}\\
\texttt{in0(i,j)=part(in,[i,j])}\\
\texttt{out0(i,j)=map(in0(i,j),mul(*,kernel(i,j)),[i,j])}\\
\texttt{elem(i,j)=reduce(out0(i,j),add,0)}\\
\texttt{conv(in,kernel):=map(in,elem(i,j),[i,j])}\\
\texttt{\#batch\_norm}\\
\texttt{mean=reduce(in,add,0)}\\
\texttt{stdvar=sqrt(reduce(map(in,sub(*,mean)),add,0))}\\
\texttt{bn(in,mean,stdvar):=map(in,sub(*,mean)/stdvar)}\\
\texttt{\#relu}\\
\texttt{relu(in):=map(in,max(*,0))}\\
\texttt{\#basic\_block}\\
\texttt{residual=in}\\
\texttt{out=conv.bn.relu.conv.bn(in)}\\
\texttt{forward(in):=relu(out+residual)}\\

\subsubsection{Representation for clustering}

We select K-means in classical clustering algorithms~\cite{krishna1999genetic} as the example. Then, we choose the code from sklearn.cluster, and the following is the very simplified version. We ignore the while-loop as we consider the computation rather than the control flow.

\textbf{Algorithm 2.1: K-means in clustering algorithm in Python}
\begin{lstlisting}[language=Python]
class KMeans(object):
 #distance
 def distance(self, i, center):
 diff = self.data[i] - center
 return np.sum(np.power(diff, 2))**0.5
 #center
 def center(self, cluster):
 cluster = np.array(cluster)
 return ...
 #main
 def cluster(self):
 changed = True
 while changed: #while-loop
  self.clusters = []
  for i in range(self.k):
  self.clusters.append([])
  for i in range(self.capacity):
  min_distance = sys.maxsize
  center = -1
    for j in range(self.k):
     distance = self.distance(i,self.centers[j])
     if min_distance > distance:
      min_distance = distance
      center = j
    self.clusters[center].append(self.data[i])
    newCenters = []
    for cluster in self.clusters:
    newCenters.append(selfcenter(cluster).tolist())
    self.centers = np.array(newCenters)
\end{lstlisting}

\textbf{Algorithm 2.2: K-means in clustering algorithm in ToLang}
\texttt{\#distance}\\
\texttt{coordinate=map(data,sqrt(pow(*-center,2)))}\\
\texttt{dist(data,center):=reduce(coordinate,add,0)}\\
\texttt{\#center}\\
\texttt{n=|data|}\\
\texttt{\textbf{tol} centers=[k]}\\
\texttt{centers(data):=map(centers,rand(n),j)}\\
\texttt{\#main}\\
\texttt{\#loop}\\
\texttt{\textbf{tol} newcenters=[k]}\\
\texttt{newcenters=map(reduce(dist(data[i],centers[j]),min,+Inf,i),j)}\\
\texttt{cluster=newcenters}\\

\begin{table}[ht!]
\centering
\resizebox{\linewidth}{!}{
\begin{tabular}{llll}
\hline
Operator                                  & Representation                                                                                                                                                                      & FLOPs                                    & EOPs                                     \\\hline
ArgMax(t)$\mapsto$axes                    & \texttt{\tabincell{l}{reduce(t(i),max,0,iter)/t[1]\\result=iter}}                                                                                                                                   & vol(t)                                   & vol(t)                                   \\\hline
ArgMin(t)$\mapsto$axes                    & \texttt{\tabincell{l}{reduce(t(i),min,0,iter)/t[1]\\result=iter}}                                                           & vol(t)                                   & vol(t)                                   \\\hline
AveragePool(t,k,p,s)$\mapsto$t            & \texttt{\tabincell{l}{\textbf{tol}   result=[(|in|[1]+2*p-k)/s+1]\\in\_pad=[\textbf{tol}[p,|in|],in,\textbf{tol}[p,|in|]]\\vec(i)=part(in\_pad,[(i-1)*s+1,i*s])\\elem(i)=reduce(vec(i),add,0)/s\\result=map(result,elem(i),i)}}                                                                             & \tabincell{l}{$\Pi$(|t|[axes]+2*p\\-k)/s+1)*k}        & \tabincell{l}{$\Pi$(|t|[axes]+2*p\\-k)/s+1)*k}        \\\hline
BatchNormalization(t,mean,var)$\mapsto$t  & \texttt{\tabincell{l}{map(t,(*-mean)/var)}}                                                                                                                      & 2*vol(t)                                 & 3*vol(t)                                 \\\hline
Celu(t,alpha)$\mapsto$t                   & \texttt{\tabincell{l}{map(t,max(*,0)+min(0,alpha*exp(*/alpha)-1))}}                                                                                              & -                                        & 6*vol(t)                                 \\\hline
Clip(t,mi,ma)$\mapsto$t                   & \texttt{\tabincell{l}{map(t,max(mi,min(*,ma)))}}                                                                                                                 & 2*vol(t)                                 & 2*vol(t)                                 \\\hline
Concat([t,...])$\mapsto$t             & \texttt{\tabincell{l}{result=[t,...]}}                                                                                                                       & 0                                        & 0                                        \\\hline
Conv(t,k,p,s)$\mapsto$t                   & \texttt{\tabincell{l}{\textbf{tol}   result=[(|in|[1]+2*p-k)/s+1]\\in\_pad=[\textbf{tol}[p,|in|],in,\textbf{tol}[p,|in|]]\\in0(i)=part(in\_pad,[(i-1)*s+1,i*s])\\out0(i)=map(in0(i),mul(*,k(i)),[i])\\elem(i)=reduce(out0(i),add,0)\\result=map(result,elem(i),[i])}} & \tabincell{l}{$\Pi$((|t|[axes]+2*p\\-|k|)/s+1)*2*|k|} & \tabincell{l}{$\Pi$((|t|[axes]+2*p\\-|k|)/s+1)*2*|k|} \\\hline
ConvTranspose(t,shape,p,s)$\mapsto$t      & \texttt{\tabincell{l}{\textbf{tol}   result=shape\\in\_pad=[\textbf{tol}[p,|in|],in,\textbf{tol}[p,|in|]]\\in0(i)=part(in\_pad,[(i-1)*s+1,i*s])\\result=map(result,in0(i),[i])}}                                                                                                                                                              & \tabincell{l}{$\Pi$((|t|[axes]+2*p\\-|k|)/s+1)*2*|k|} & \tabincell{l}{$\Pi$((|t|[axes]+2*p\\-|k|)/s+1)*2*|k|} \\\hline
CumSum(t,axes)$\mapsto$t                  & \texttt{\tabincell{l}{t0=reshape(t,[axes,-1])\\result=reshape(reduce(t0,add,0),|t|)}}                                          & vol(t)                                   & vol(t)                                   \\\hline
DepthToSpace(t,blocksize,shape)$\mapsto$t & \texttt{\tabincell{l}{reshape(t,[shape,blocksize])}}                                                                                                         & 0                                        & 0                                        \\\hline
Dropout(t,ratio)$\mapsto$t                & \texttt{\tabincell{l}{map(t,mul(*,rand*ratio))}}                                                                                                                 & vol(t)                                   & vol(t)                                   \\\hline
Elu(t,alpha)$\mapsto$t                    & \texttt{\tabincell{l}{map(t,max(*,0)+min(0,alpha*exp(*)-1))}}                                                                                                    & -                                        & 5*vol(t)                                 \\\hline
Expand(t,shape)$\mapsto$t                 & \texttt{\tabincell{l}{reshape(t,shape)}}                                                                                                                         & 0                                        & 0                                        \\\hline
EyeLike(k)$\mapsto$t                      & \texttt{\tabincell{l}{\textbf{tol}   result=[k,k]\\result=map(result,eq(iter[1],iter[2]))}}           & 0                                        & 0                                        \\\hline
Flatten(t)$\mapsto$t                      & \texttt{\tabincell{l}{reshape(t,vol(t))}}                                                                                                                        & 0                                        & 0                                        \\\hline
Gather(t,axes)$\mapsto$t                  & \texttt{\tabincell{l}{map(t,part(*,axes))}}                                                                                                                      & 0                                        & 0                                        \\\hline
Gemm(t1,t2)$\mapsto$t                     & \texttt{\tabincell{l}{\textbf{tol}   result=[|a|[1],|b|[2]]\\vec(i,j)=map(a[i],mul(*,b[*,j]))\\elem(i,j)=reduce(vec(i,j),add,0)\\result=map(result,elem(i,j),[i,j])}}                                                                                                                                                                                                       & 2*vol(t1)*|t2|[2]                    & 2*vol(t1)*|t2|[2]                    \\\hline
Hardmax(t,axes)$\mapsto$t                 & \texttt{\tabincell{l}{t0=reshape(t,[axes,-1])\\result=reshape(reduce(t0,max,-Inf),|t|)}}                                       & -                                        & 1                                        \\\hline
HardSigmoid(t,axes)$\mapsto$t             & \texttt{\tabincell{l}{t0=reshape(t,[axes,-1])\\result=reshape(reduce(t0,max(0,min(1,\\(1+*)/2))),|t|)}}                          & -                                        & 4*vol(t)                                 \\\hline
HardSwish(t,axes)$\mapsto$t               & \texttt{\tabincell{l}{map(t,min(0,min(*,mul(*,(*+3)/6))))}}                                                                                                      & -                                        & 5*vol(t)                                 \\\hline
Identity(t)$\mapsto$t                     & \texttt{\tabincell{l}{result=t}}                                                                                                                                 & 0                                        & 0                                        \\\hline
LeakyRelu(t,alpha)$\mapsto$t              & \texttt{\tabincell{l}{map(t,max(*,mul(alpha,*)))}}                                                                                                               & -                                        & 2*vol(t)                                 \\\hline
LogSoftmax(t,axes)$\mapsto$t              & \texttt{\tabincell{l}{t0=reshape(t,[axes,-1])\\result=reshape(map(t0,*/reduce(map(*,\\exp),add,0),|t|))\\result=map(t1,log)}}                                                                                      & -                                        & 2*vol(t)+3                               \\\hline
LpNormalization(t,p,axes)$\mapsto$t       & \texttt{\tabincell{l}{t0=reshape(t,[axes,-1])\\result=reshape(reduce(map(t,pow(*,p)),\\add,0),|t|)}}                             & 3*vol(t)                                 & 3*vol(t)                                 \\\hline
LpPool(t,k,p,s)$\mapsto$t                 & \texttt{\tabincell{l}{\textbf{tol}   result=[(|in|[1]+2*p-k)/s+1]\\in\_pad=[\textbf{tol}[p,|in|],in,\textbf{tol}[p,|in|]]\\vec(i)=part(in\_pad,[(i-1)*s+1,i*s])\\elem(i)=reduce(map(t,pow(*,p)),add,0)\\result=map(result,elem(i),i)}}                                                                      & -                                        & \tabincell{l}{$\Pi$(|t|[axes]+2*p-k)\\/s+2)*2*k+1}    \\\hline
MatMul(t1,t2)$\mapsto$t                   & \texttt{\tabincell{l}{\textbf{tol}   result=[|a|[1],|b|[2]]\\vec(i,j)=map(a[i],mul(*,b[*,j]))\\elem(i,j)=reduce(vec(i,j),add,0)\\result=map(result,elem(i,j),[i,j])}}                                                                                                                                                                                                       & 2*vol(t1)*|t2|[2]                    & 2*vol(t1)*|t2|[2]                    \\\hline
MaxPool(t,k,p,s)$\mapsto$t                & \texttt{\tabincell{l}{\textbf{tol}   result=[(|in|[1]+2*p-k)/s+1]\\in\_pad=[\textbf{tol}[p,|in|],in,\textbf{tol}[p,|in|]]\\vec(i)=part(in\_pad,[(i-1)*s+1,i*s])\\elem(i)=reduce(vec(i),max,0)\\result=map(result,elem(i),i)}}                                                                               & \tabincell{l}{$\Pi$(|t|[axes]+2*p\\-k)/s+1)*k}        & \tabincell{l}{$\Pi$(|t|[axes]+2*p\\-k)/s+1)*k}        \\\hline
\end{tabular}}
\caption{(Part 1/2) ONNX (Open Neural Network Exchange) is an open format built to represent machine learning models. It covers sophisticated computations on tensors and other data formats. This table summarizes how to present the ONNX operators in ToLang. Column 1 is ONNX operators with their parameter formats. Column 2 is a ToLang representation in ToL primitive operators. Column 3 is the FLOPs of the state-of-the-practice algorithms. Column 4 is EOPs derived from the ToL counting rules. Note that ``*'' refers to an iterator.}
\label{tab31}
\end{table}

\begin{table}[ht!]
\centering
\resizebox{\linewidth}{!}{
\begin{tabular}{llll}
\hline
Operator                                  & Representation                                                                                                                                                                      & FLOPs                                    & EOPs                                     \\\hline
MaxUnpool(t,shape,p,s)$\mapsto$t          & \texttt{\tabincell{l}{\textbf{tol}   result=shape\\in\_pad=[\textbf{tol}[p,|in|],in,\textbf{tol}[p,|in|]]\\vec(i)=[in[i],…(shape[1]/s),in[i]]\\result=map(result,vec(i),i)}}                                                                                                                                                          & \tabincell{l}{$\Pi$(|t|[axes]+2*p\\-k)/s+1)*k}        & \tabincell{l}{$\Pi$(|t|[axes]+2*p\\-k)/s+1)*k}        \\\hline
Mean(t)$\mapsto$t                         & \texttt{\tabincell{l}{reduce(t(i),add,0)/|t|}}                                                                                                                   & vol(t)                                   & vol(t)                                   \\\hline
OneHot(n,indices,depth,values)$\mapsto$t  & \texttt{\tabincell{l}{\textbf{tol}   result=[n,depth]\\result=map(\textbf{tol}[n],eq(*,iter))}}                                                                     & -                                        & 0                                        \\\hline
Pad(t,p)$\mapsto$t                        & \texttt{\tabincell{l}{\textbf{tol}   result=[\textbf{tol}[p,|t|],in,\textbf{tol}[p,|t|]]}}        & 0                                        & 0                                        \\\hline
PRelu(t,slope)$\mapsto$t                  & \texttt{\tabincell{l}{map(t,max(*,mul(slope,*)))}}                                                                                                               & -                                        & 2*vol(t)                                 \\\hline
Range(start,limit,delta)$\mapsto$t        & \texttt{\tabincell{l}{\textbf{tol}   in=[limit-start]\\result=map(in,iter*delta)}}                            & vol(t)                                   & vol(t)                                   \\\hline
ReduceL1(t,axes)$\mapsto$t                & \texttt{\tabincell{l}{t0=reshape(t,[axes,-1])\\result=reshape(reduce(t0,add,0),|t|)}}                                          & 2*vol(t)                                 & 2*vol(t)                                 \\\hline
ReduceL2(t,axes)$\mapsto$t                & \texttt{\tabincell{l}{t0=reshape(t,[axes,-1])\\result=reshape(map(reduce(map(t0,pow(*,2)),\\add,0),sqrt),|t|)}}                  & 2*vol(t)                                 & 2*vol(t)                                 \\\hline
ReduceLogSum(t,axes)$\mapsto$t            & \texttt{\tabincell{l}{t0=reshape(t,[axes,-1])\\result=reshape(map(reduce(t0,add,0),log),|t|)}}                                 & 2*vol(t)                                 & 2*vol(t)                                 \\\hline
ReduceLogSumExp(t,axes)$\mapsto$t         & \texttt{\tabincell{l}{t0=reshape(t,[axes,-1])\\result=reshape(map(reduce(map(t0,exp),\\add,0),log),|t|)}}                        & 3*vol(t)                                 & 3*vol(t)                                 \\\hline
ReduceMax(t,axes)$\mapsto$t               & \texttt{\tabincell{l}{t0=reshape(t,[axes,-1])\\result=reshape(reduce(t0,max,-Inf),|t|)}}                                       & vol(t)                                   & vol(t)                                   \\\hline
ReduceMean(t,axes)$\mapsto$t              & \texttt{\tabincell{l}{t0=reshape(t,[axes,-1])\\result=reshape(map(reduce(t0,add,0),\\*/shape(*)),|t|)}}                          & vol(t)                                   & vol(t)                                   \\\hline
ReduceMin(t,axes)$\mapsto$t               & \texttt{\tabincell{l}{t0=reshape(t,[axes,-1])\\result=reshape(reduce(t0,min,+Inf),|t|)}}                                       & vol(t)                                   & vol(t)                                   \\\hline
ReduceProd(t,axes)$\mapsto$t              & \texttt{\tabincell{l}{t0=reshape(t,[axes,-1])\\result=reshape(reduce(t0,mul,1),|t|)}}                                          & vol(t)                                   & vol(t)                                   \\\hline
ReduceSum(t,axes)$\mapsto$t               & \texttt{\tabincell{l}{t0=reshape(t,[axes,-1])\\result=reshape(reduce(t0,add,0),|t|)}}                                          & vol(t)                                   & vol(t)                                   \\\hline
ReduceSumSquare(t,axes)$\mapsto$t         & \texttt{\tabincell{l}{t0=reshape(t,[axes,-1])\\result=reshape(reduce(map(t0,pow(*,2)),\\add,0),|t|)}}                            & 2*vol(t)                                 & 2*vol(t)                                 \\\hline
Relu(t)$\mapsto$t                         & \texttt{\tabincell{l}{map(t,max(*,0))}}                                                                                                                          & -                                        & vol(t)                                   \\\hline
Reshape(t,sha)$\mapsto$t                  & \texttt{\tabincell{l}{reshape(t,sha)}}                                                                                                                           & 0                                        & 0                                        \\\hline
Selu(t,alpha,gamma)$\mapsto$t             & \texttt{\tabincell{l}{map(t,gamma*max(*,0)+min(0,alpha*exp(*)-1))}}                                                                                              & -                                        & 6*vol(t)                                 \\\hline
Shape(t)$\mapsto$t                             & \texttt{\tabincell{l}{|t|}}                                                                                                                                      & 0                                        & 0                                        \\\hline
Sigmoid(t)$\mapsto$t                      & \texttt{\tabincell{l}{map(t,1/(1+exp(-*)))}}                                                                                                                     & -                                        & 4*vol(t)                                 \\\hline
Size(t)$\mapsto$ui                        & \texttt{\tabincell{l}{vol(t)}}                                                                                                                                   & 0                                        & 0                                        \\\hline
Slice(t,starts,ends,axes,steps)$\mapsto$t & \texttt{\tabincell{l}{t0=reshape(t,[axes,-1])\\result=map(t0,part(*,map(\textbf{tol}[(ends-start)\\/steps],iter*steps)))}}                                                                                                  & 0                                        & 0                                        \\\hline
Softmax(t,axes)$\mapsto$t                 & \texttt{\tabincell{l}{t0=reshape(t,[axes,-1])\\t1=map(t0,*/reduce(map(*,exp),add,0))\\result=reshape(t1,|t|)}}                                                                                                   & -                                        & 2*vol(t)+2                               \\\hline
Softplus(t)$\mapsto$t                     & \texttt{\tabincell{l}{map(t,log(1+exp(*)))}}                                                                                                                     & -                                        & 3*vol(t)                                 \\\hline
Softsign(t)$\mapsto$t                     & \texttt{\tabincell{l}{map(t,*/(*,sgn(x)))}}                                                                                                                      & -                                        & 3*vol(t)                                 \\\hline
SpaceToDepth(t,blocksize)$\mapsto$t       & \texttt{\tabincell{l}{reshape(t,[blocksize])}}                                                                                                               & 0                                        & 0                                        \\\hline
Split(t,axes,ui)$\mapsto$[t,...]      & \texttt{\tabincell{l}{t0=reshape(t,[axes,-1])\\result=[t0[1],...,t0[|t0|]]}}                                       & 0                                        & 0                                        \\\hline
Sum([t,...])$\mapsto$t                & \texttt{\tabincell{l}{reduce([t,...],add,\textbf{tol}|t|)}}                                                                                 & vol(t)*n                                 & vol(t)*n                                 \\\hline
Swish(t)$\mapsto$t                        & \texttt{\tabincell{l}{map(t,mul(*,1/(1+exp(-*))))}}                                                                                                              & -                                        & 5*vol(t)                                 \\\hline
ThresholdedRelu(t,th)$\mapsto$t           & \texttt{\tabincell{l}{map(t,mul(gt(*,th),*))}}                                                                                                                   & -                                        & 2*vol(t)                                 \\\hline
Transpose(t,axes)$\mapsto$t               & \texttt{\tabincell{l}{reshape(t,swap(|t|,axes))}}    & 0                                        & 0        \\\hline
\end{tabular}}
\caption{(Part 2/2) ONNX (Open Neural Network Exchange) is an open format built to represent machine learning models. It covers sophisticated computations on tensors and other data formats. This table summarizes how to present the ONNX operators in ToLang. Column 1 is ONNX operators with their parameter formats. Column 2 is a ToLang representation in ToL primitive operators. Column 3 is the FLOPs of the state-of-the-practice algorithms. Column 4 is EOPs derived from the ToL counting rules. Note that ``*'' refers to an iterator.}
\label{tab32}
\end{table}

%% file: 6.Performance.tex
\section{ToL has a built-in practical performance indicator}

This section shows ToL has a built-in practical performance indicator -- elementary operation counts (EOPs) -- which is consistent with the widely-used FLOPs~\cite{2010The} metric.



\subsection{The definition of elementary operation count (EOPs) and counting rules}\label{EOPs_count_rule}

ToL is born with a practical performance indicator -- elementary operation counts (EOPs), counting how many basic elementary functions are involved in an algorithm. From the definitions in Section~\ref{Section:ToLang}, we know that the EOPs of an algorithm are definite as they are compositions of ToL operators whose  EOPs are also definite according to the rules.  The detailed counting rules are as follows.



The EOPs of a function is determined by the number of the basic elementary functions that compose the function. The EOPs of a function is equal to the sum of the EOPs of its basic elementary functions. We consider that the EOPs of each basic elementary function (see Table~\ref{tab2}) is one.

\textbf{Example 13}: For variable $x$ and vector $v$ sized $n$, $f(x)=a*x+b$, $g(v)=sqrt(v_1^2+v_2^2+\cdots+v_n^2)$. The operation counts are computed as follows.\\ EOPs$(f(x))=$EOPs$(*)+$EOPs$(+)$=\texttt{2}\\ EOPs$(g(v))=n*$EOPs$(*)+(n-1)*$EOPs$(+)+$EOPs$(sqrt)$=\texttt{2*n}\\
This result (\texttt{2} and \texttt{2*n}) is consistent with FLOPs count for $f(x)$ and $f(x)$ ($2$ FLOPs and $2n$ FLOPs).

The EOPs of a ToL operator depends on its variable data quantity (\texttt{vol(<var>)*capacity(<var>)}) and the basic elementary functions involved. We define the counting rules of each ToL operator as follows.

\begin{itemize}
\item EOPs(\texttt{single\_var})=\texttt{0}
\item EOPs(\texttt{f.g.h....(<var>)})=EOPs(\texttt{f})+EOPs(\texttt{g})+EOPs(\texttt{h})+...+EOPs(\texttt{<var>})\\//rule for composite function
\item EOPs(\texttt{shape})=EOPs(\texttt{<var>})
\item EOPs(\texttt{dim})=EOPs(\texttt{<var>})
\item EOPs(\texttt{capacity})=EOPs(\texttt{<var>})
\item EOPs(\texttt{vol})=EOPs(\texttt{<var>})
\item EOPs(\texttt{space})=EOPs(\texttt{<var>})
\item EOPs(\texttt{convert})=EOPs(\texttt{<var>})
\item EOPs(\texttt{part})=EOPs(\texttt{<var>})
\item EOPs(\texttt{swap})=EOPs(\texttt{<var>})
\item EOPs(\texttt{reshape})=EOPs(\texttt{<var>})
\item EOPs(\texttt{tile})=EOPs(\texttt{<var>})
\item EOPs(\texttt{map(<var>,<func>,<iter>)})=\texttt{vol(<var>)*capacity(<var>)*}EOPs(\texttt{<func>})+EOPs(\\\texttt{<var>})
\item EOPs(\texttt{reduce(<var>,<func>,<var\_0>,<iter>)})=\texttt{vol(<var>)*capacity(<var>)*}EOPs(\\\texttt{<func>})+EOPs(\texttt{<var>})+EOPs(\texttt{<var\_0>})
\end{itemize}


We present matrix multiplication as an example to show how to count the EOPs of an algorithm in detail.

\textbf{Example 14}: Consider EOPs of matrix multiplication represented as \texttt{result=}MatMul\texttt{(a,b)}, shown in Section~\ref{section521}. Suppose \texttt{a} and \texttt{b} are matrices shaped \texttt{[n,n]}.\\
EOPs(\texttt{result})\\
=EOPs(\texttt{map(result,elem(i,j),[i,j])})\\
=EOPs(\texttt{result})+\texttt{|result|*}EOPs(\texttt{elem(i,j)})\\
=\texttt{n*n*}EOPs(\texttt{reduce(vec(i,j),add,0)})\\
=\texttt{n*n*}(EOPs(\texttt{vec(i,j)})+\texttt{|vec(i,j)|*}EOPs(\texttt{add}))\\
=\texttt{n*n*}(EOPs(\texttt{map(a[i],mul(*,b[*,j]))}+\texttt{n*1})\\
=\texttt{n*n*}(EOPs(\texttt{a[i]})+\texttt{|a[i]|}*EOPs(\texttt{mul})+\texttt{n})\\
=\texttt{n*n*(n+n)}\\
=\texttt{2*n*n*n}

This result (\texttt{2*n*n*n}) is consistent with the FLOPs count for matrix multiplication ($2n^3$ FLOPs).


\subsection{Evaluations on the ONNX operators}

This subsection uses the ONNX~\cite{onnx} operators as an example to show that the EOPs metric is consistent with the FLOPs in a complex application scenario like AI or big data.


According to the counting rules defined in the last section, we compute the EOPs and the FLOPs (the third and fourth columns) of ONNX operators in Table~\ref{tab31} and \ref{tab32}. We obtain the FLOPs after analyzing the state-of-the-practice implementations of the ONNX operators and calculate the EOPs according to the counting rules defined in the last section. The comparisons show that EOPs are consistent with  FLOPs within a minimal error range. This evaluation further shows that ToLang has a built-in realistic performance indicator.

%% file: 7.Conclusion.tex
\section{Conclusion}
This article presented a unified computation model with a generalized expression ability: ToL. ToL includes a unified data abstraction -- Tensor of List, and a unified computation model based on Tensor of List. ToL introduces five atomic computations that can represent any elementary computation by finite compositions, ensured with strict formal proof. Based on ToL, we designed a pure-functional language -- ToLang. ToLang provides a concise set of primitive operators that can be used to program complex big data and AI algorithms. The evaluations showed ToL has generalized expression ability and a built-in performance indicator, born with a strictly defined computation metric -- elementary operation count (EOPs), consistent with FLOPs within a small error range.



